\documentclass[12pt]{article}
\begin{document}
\begin{center}
{\bf Proper fluctuations associated with quantum tunneling in
field theory}

\vspace{0.8cm} { Michael Maziashvili }

\vspace{0.5cm}\baselineskip=14pt

\vspace{0.5cm} \baselineskip=14pt {\it Department of Theoretical
Physics, Tbilisi State University, 0128 Tbilisi, Georgia}
\vspace{0.1cm} maziashvili@ictsu.tsu.edu.ge
\end{center}
\vspace{0.5cm}

\begin{abstract}
It is shown that, to the lowest order in $\hbar,$ the particle
production related to the tunneling that leads to the false vacuum
decay is described by the orthogonal part of fluctuation field
with respect to the bounce solution. As a simple example the
spatially homogeneous tunneling is considered in order to
illustrate the consequences coming from such a restriction of the
fluctuation field.
\end{abstract}
\section{Introduction}
Coupling of tunneling field $\phi$ to the fluctuation one leads to
the particle production effect during the tunneling process. To
make this more precise, consider the potential $U(\phi)$ shown in
Fig.1. There are two minima, $\phi_{\pm},$ both of which
correspond to classically stable homogeneous field configurations.
The false vacuum corresponds to $\phi=\phi_{+}$ for which we
assume $U(\phi_{+})=0$. In the WKB approximation the tunneling
probability per unit volume per unit time of such a state is of
the form
\begin{equation}\label{pvpt}P/VT=Ae^{-B/\hbar}(1+O(\hbar)).\end{equation} Coleman \cite{Col}
showed that to find $B$ one has to solve the field equations of
the theory in Euclidean space-time subjected to the boundary
condition that the field asymptotically approaches its false
vacuum value. In addition to the trivial solution $\phi=\phi_{+}$
there is a time-reversal invariant zero-energy solution, $\phi_b$,
referred to as a bounce one and $B$ is the Euclidean action
evaluated at the bounce. The prefactor $A$ contains the first
quantum correction to the tunneling process. The problem of
particle creation during the tunneling process, leading to the
decay of false vacuum in quantum field theory,  was first studied
in \cite{Rub}. The further development of the general formalism
describing this process is given in \cite{Vach-Vil,TSY,YTS,HSTY}.
Recently the particle spectrum in the thin-wall approximation has
been considered in \cite{Ma}. It was shown that in this case the
particle production is strongly suppressed. The purpose of this
brief paper is to point out the existence of a constraint on the
fluctuation field in the tunneling process. This constraint arises
due to fact that the fluctuation field associated with the
tunneling that gives rise to the particle production is the
transverse part of total fluctuation field with respect to the
bounce solution. This is in fact implicit in \cite{Bit-Ch}.
Namely, the one-loop functional determinant obtained in this
article is explicitly written as a functional integral over
transverse part of fluctuation field. Nevertheless, this fact has
not received attention so far in consideration of particle
production during the tunneling process in field theory. We
briefly notice here that for evaluating the preexponential factor,
$A$, Callan and Coleman \cite{Call-Col} used an Euclidean version
of Hamilton's action in the path-integral formalism. But this
approach has the serious drawbacks considered by Patrascioiu
\cite{Pat}. In Sec.2 we specify the fluctuation field associated
with the tunneling process and, in this context, consider an
example of spatially homogeneous tunneling.. The conclusions are
briefly summarized in Sec.3.
\section{Proper fluctuations associated with the tunneling}
 At first we
demonstrate that in order to recover the one-loop determinant
expression of \cite{Bit-Ch} in the path integral formalism one has
to use the Jacobi type action. Because this Jacobi type action
corresponds to the Euclidean Hamilton's action in much the same
way that connects Jacobi's action to Hamilton's one, we refer to
it as the Euclidean Jacobi's action. The Euclidean Jacobi's action
has the form
\begin{equation}\label{Eac}J_E[\phi]=\int
\limits_{-\infty}\limits^{\infty}\!\!d\sigma\sqrt{2V[\phi]\int\!
d^3y\dot{\phi}^2 },\label{Ev}\end{equation} where $V[\phi]=\int
d^3y\left(\frac{(\vec{\nabla}\phi)^2}{2}+U(\phi)\right)$ and
$\dot{\phi}\equiv\frac{d\phi}{d\sigma}.$ It is obtained from the
Jacobi's action $J$ formally as $\sigma\rightarrow
-\imath\sigma,~J_E\rightarrow \imath J.$ We want to estimate the
following functional integral in the semiclassical(small-$\hbar$)
limit.
\begin{equation} P\equiv\int\limits_{\phi (-\infty)=\phi_f}^{\phi (\infty)=\phi_f}
\!\![D\phi]_{FP}\,e^{-\frac{1}{\hbar}\int
_{-\infty}\limits^{\infty}\!\!d\sigma \sqrt{2V[\phi]\int\!
d^3x\dot{\phi}^2 }},\label{pi}\end{equation} where $[D\phi]_{FP}$
is a Faddeev-Popov measure and the integration is over all
functions $\phi (\sigma,\vec{x})$ obeying the boundary conditions
$\phi (\pm\infty,\vec{x})=\phi_{+}.$ As an essential point for our
discussion we want to emphasize that the action (\ref{Ev}) is
invariant under the reparameterizations of the configuration
space-path that preserve the end point values of the parameter.
That is, (\ref{Ev}) is invariant under the replacements $\sigma
\rightarrow f(\sigma)$ and $\phi(\sigma
,\vec{x})\rightarrow\bar{\phi}(f(\sigma), \vec{x})$ with
$f(\pm\infty)=\pm\infty$. Their infinitesimal form is $\sigma
\rightarrow\sigma+\epsilon(\sigma)$ and $\phi\rightarrow
\phi+\epsilon\dot{\phi}$ where $\epsilon(\pm\infty)=0$. Now it is
obvious that the proper fluctuations for action (\ref{Ev}) are
transverse ones $\delta\phi_{\perp}(\vec{x})=\int d^3y\Pi_{\perp}
(\vec{x},\vec{y})\delta\phi (\vec{y}),$ where $\Pi_{\perp}
(\vec{x},\vec{y})=\delta (\vec{x}-\vec{y})-\dot{\phi}(\vec{x})
\dot{\phi}(\vec{y})/\int d^3z\dot{\phi}^2$ is the projection
operator onto the subspace of configuration space that is
orthogonal to the configuration space-path $\phi
(\sigma,\vec{x})$, while the longitudinal fluctuations reproduce a
gauge transformation.  For $\hbar\rightarrow 0$ the path integral
is dominated by the contribution of stationary point of action
(\ref{Ev}). In imaginary time parameterization, $\sigma (\tau)$,
obtained by using the Euclidean zero-energy condition
\begin{equation}\int\!
d^3y\frac{1}{2}\left(\frac{d\phi}{d\tau}\right)^2-V[\phi]=0,\end{equation}
this yields the bounce solution $\phi_{b}$. Since we are
interested in evaluation of that functional integral for
$\hbar\rightarrow 0$ it is sufficient to restrict ourselves to the
consideration of small fluctuations around the bounce solution.
Thus, we can use the Faddeev-Popov method \cite{Fadd-Pop}
immediately for the infinitesimal gauge transformation. Taking
into account that the longitudinal fluctuations reproduce a gauge
transformation then following to the Faddeev-Popov procedure one
has to integrate the quadratic action over transverse ones, which
are proper fluctuations for the action (\ref{Eac}). It may be done
by splitting of the transverse and longitudunal fluctuations,
$D\delta\phi =JD\delta\phi _{\perp}D\delta\phi _{\parallel},$
where $J$ is Jacobian associated with this transformation and
integrating the path-integral over (proper) transverse
fluctuations with the Faddeev-Popov measure,
$[D\delta\phi]_{FP}\equiv JD\delta\phi _{\perp}$. The integration
over the gauge degrees of freedom, longitudinal fluctuations, is
dropped. Expanding about the bounce solution to quadratic order in
fluctuation field the functional integral (\ref{pi}) becomes
\begin{equation}\label{Amp}P=e^{-\frac{1}{\hbar}J_E[\phi_b]}\int\!
JD\delta\phi _\perp\exp\left[-\frac{1}{2\hbar}\int\!d^4x\delta\phi
_{\perp}\left(-\frac{\partial ^2}{\partial\tau
^2}-\triangle+U''(\phi_b)\right)\delta\phi
_{\perp}\right],\end{equation} with the boundary conditions
$\delta\phi_{\perp}(\pm\infty,\vec{x})=0.$ Here
$\delta\phi_{\perp}(\vec{x})=\int d^3y\Pi_{\perp}^b
(\vec{x},\vec{y})\delta\phi (\vec{y})$ and $\Pi_{\perp}^b
(\vec{x},\vec{y})$ denotes orthogonal projection onto the bounce
solution. The integration over the zero modes associated with the
space and time translations gives an explicit $VT$ factor in
computing the vacuum tunneling amplitude. So, in the appropriate
limit our result recovers those obtained in \cite{Bit-Ch}. On the
other hand the stationary zero-energy Schr\"{o}dinger equation
describing this tunneling event may be obtained by the
quantization of the zero-energy Jacobi's action,
\begin{equation}\label{Jacobi}J[\phi]=\int\limits _{-\infty}\limits^{\infty}d\sigma
\sqrt{-2V[\phi]\int\! d^3y\dot{\phi}^2 }.\end{equation} The
relationship between Jacobi's and Hamilton's action principles as
well as their quantization is considered in detail in \cite{Br-Y}.
Because this Lagrangian is homogenous of degree one in the
$\dot{\phi}(\vec{x})$, the canonical Hamiltonian vanishes
identically. This is of course a well-known feature of the
reparametrization invariant theories. Consequently the momenta
conjugate to the $\phi(\vec{x})$, $\pi
(\vec{x})=\dot{\phi}(\vec{x})\sqrt{-2V[\phi]/\int
d^3y\dot{\phi}^2},$ gives rise to a
constraint\begin{equation}\label{cons}\int\!
d^3y\frac{\pi^2(\vec{y})}{2}+V[\phi]=0.\end{equation} There are no
secondary constraints and the constraint (\ref{cons}) is then
trivially first class. Due to the Dirac's procedure this
constraint is imposed as a quantum operator equation acting on the
wave functional. In the coordinate representation
$\hat{\phi}(\vec{x})=\phi
(\vec{x}),~\hat{\pi}(\vec{x})=-\imath\hbar\frac{\delta}{\delta\phi
(\vec{x}) }$ and the wave functional $\Psi [\phi]$ depends on
$\phi (\vec{x}).$ The resulting operator-constraint equation is
just the stationary zero-energy Schr\"{o}dinger equation
\begin{equation}\label{Sch}\left(-\frac{\hbar^2}{2}\int\!
d^3y\frac{\delta^2}{\delta\phi(y)^2}+V[\phi]\right)\Psi[\phi]=0.\end{equation}
This equation is obviously obtained as a result of starting with
Jacobi's action, in which time nowhere appears but the energy is
fixed $E=0.$ According to the formalism developed by Bitar and
Chang \cite{Bit-Ch} the lowest-order WKB approximation to this
equation shows that the wave functional $\Psi$ is strongly peaked
around a one-parameter family of field configurations
$\tilde{\phi}(\sigma,\vec{x})$ which in a classically forbidden
region is determined by the stationary point of Euclidean Jacobi's
action (\ref{Eac}) and by the stationary point of Jacobi's action
(\ref{Jacobi}) in the classically allowed region. As it was
discussed above, the proper fluctuation field about the
$\tilde{\phi}$ corresponding to the (Euclidean)Jacobi's action is
the transverse part of fluctuation field with respect to this
field configuration. To take into account the effect of
fluctuations to this tunneling process one has to split the field
into the tunneling one and the fluctuation field around it
$\phi\rightarrow \phi+\delta\phi_{\perp}.$ For the combined system
of $\phi+\delta\phi_{\perp}$ one gets again a zero-energy
stationary Schr\"{o}dinger equation from the Jacobi's action
(\ref{Jacobi}). Thus one simply concludes that the basic equation
governing the fluctuation field $\delta\phi$ during the tunneling
must be supplemented by the constraint
\begin{equation}\label{con}\int
d^3x\delta\phi(\tau,\vec{x})\phi'_b/\sqrt{\tau^2+\vec{x}\,^2}=0,\end{equation}
where we have taken into account the $O(4)$ symmetry of the bounce
solution \cite{Col} and the prime denotes differentiation with
respect to $\rho=\sqrt{\tau^2+\vec{x}\,^2}$. The constraint in the
Minkowskian region is obtained from (\ref{con}) by the analytic
continuation $\tau\rightarrow it$. To see the idea let us consider
an example of spatially homogeneous tunneling considered
previously in \cite{Rub,TSY}. In this case the tunneling
configuration denoted by $\phi_b$ is spatially homogeneous,
$\phi_b(\tau)$. Following \cite{TSY} we assume $U''(\phi_b(\tau))$
to be a step function
\begin{equation} U''(\phi_b(\tau))=\left\{\begin{array}{ll} m_{+}^2 &\mbox{$\tau<\tilde{\tau},$}\\
m_{-}^2
&\mbox{$\tau>\tilde{\tau},$}\end{array}\right.\end{equation} where
$\tilde{\tau}<0$ with a large absolute value. The equation
governing the fluctuation field takes the form
\begin{equation}\label{bas}\begin{array}{ll} (\partial^2_{\tau}+\triangle+m^2_{+})\delta\phi=0 &\mbox{$\tau<\tilde{\tau},$}\\
(\partial^2_{\tau}+\triangle+m^2_{-})\delta\phi=0
&\mbox{$\tau>\tilde{\tau}.$} \end{array}\end{equation} Taking the
mode expansion $\delta\phi=(2\pi)^{-3/2}\int
v_{\vec{p}}\,(\tau)e^{i\vec{p}\,\vec{x}} d^3\!p$ one easily finds
the (unnormalized) solution to Eq.(\ref{bas}) satisfying the
vanishing boundary condition when
$\tau\rightarrow-\infty$ \begin{equation} v_{\vec{p}}(\tau)=\left\{\begin{array}{ll} e^{\omega_{+}\tau} &\mbox{$\tau<\tilde{\tau},$}\\
a_{+}(\vec{p})e^{\omega_{-}\tau}+a_{-}(\vec{p})e^{-\omega_{-}\tau}
&\mbox{$\tau>\tilde{\tau},$}\end{array}\right.\end{equation} where
$\omega_{\pm}=\sqrt{\vec{p}\,^2+m_{\pm}^2}$ and
\[a_{\pm}=\frac{1}{2\omega_{-}}(\omega_{-}\pm\omega_{+})e^{\mp(\omega_{-}\mp\omega_{+})\tilde{\tau}}.\] The spectrum of created particles has the form
\cite{TSY}
\begin{equation}\label{spect}n(p)=\frac{1}{\left(\frac{\omega_{-}+\omega_{+}}{\omega_{-}-\omega_{+}}\right)^2e^{-4\omega_{-}\tilde{\tau}}-1}.\end{equation}
Furthermore, the constraint (\ref{con}), that the fluctuation
field must be orthogonal to the $\phi_b$, implies now that the
integral of $\delta\phi$ over all space vanishes,
\begin{equation}v_{{\bf 0}}(\tau)\propto\int d^3x\delta\phi =0.\end{equation}
So, we arrive at the result that the spatially homogeneous
tunneling does not allow the particle production with zero
momentum. Correspondingly, for $\vec{p}={\bf 0}$ one can not use
the Eq.(\ref{spect}), which remains valid for $p>0$.
\section{Summary}
We have demonstrated  that the proper fluctuation field associated
with the tunneling process is the transverse part of fluctuation
field with respect to the bounce solution. This statement is quite
natural since the stationary Schr\"{o}dinger equation needed to
describe the tunneling phenomenon is obtained by the quantization
of Jacobi's action for which the longitudinal part of fluctuation
field reproduces the gauge transformation. Correspondingly, we
point out that the general formalism describing the particle
production in the tunneling process must be supplemented by the
constraint (\ref{con}). The use of this constraint is demonstrated
in the case of a spatially homogenous tunneling. Roughly speaking,
one must subtract the number of particles created by the
longitudinal part of fluctuation field from the particle spectrum
obtained by the total fluctuation field. Of course such a
subtraction will not affect the conclusion made in \cite{Ma} that
in general the particle production in the thin-wall approximation
is exponentially suppressed. It is of interest to explicate the
consequences coming from the constraint (\ref{con}) in the
realistic cases.

\section*{Acknowledgments}
The author is greatly indebted to his advisor in TSU Professor
A.\,Khelashvili and to Professor G.\,Lavrelashvili for helpful
conversations.

\end{document}